\pdfoutput=1
\documentclass[journal,onecolumn,12pt,twoside]{IEEEtranTCOM}
\usepackage{etex}

\normalsize

\usepackage{cite}
\usepackage{setspace}

\usepackage{graphicx}
\usepackage{color}
\usepackage{pst-sigsys}
\ifCLASSINFOpdf
\else
\fi

\usepackage[cmex10]{amsmath}
\usepackage{amsmath,dsfont}
\usepackage{verbatim}
\usepackage{amssymb}
\usepackage{amsthm}
\usepackage{nccmath}
\usepackage{enumerate}
  \usepackage{pgfplots}
  \usepackage[normalem]{ulem}
  \usepackage{cases}
  \usepackage{bbm}  
\pgfplotsset{plot coordinates/math parser=false}
\newtheorem{mydef}{Definition}

\newtheorem{remark}{Remark}
\newcommand{\Rmnum}[1]{\MakeUppercase{\romannumeral #1}}

\newcommand{\nRepi}[1]{\ensuremath{M_{#1}}}
\newcommand{\vecnRep}{\ensuremath{\underline{M}}}
\newcommand{\indi}[1]{\ensuremath{\mathbbm{1}_{#1}}}

\newcommand{\Nretransmit}{\ensuremath{N_{0}}}
\newcommand{\ENretransmit}{\ensuremath{\mathbb{E}\Nretransmit}}

\newcommand{\astar}{\ensuremath{c^{\dagger}}}
\newcommand{\bstar}{\ensuremath{d^{\dagger}}}
\newcommand{\cstar}{\ensuremath{a^{\dagger}}}
\newcommand{\mydstar}{\ensuremath{b^{\dagger}}}

\newcommand{\omegaparam}{\ensuremath{\theta}}
\newcommand{\gammaparam}{\ensuremath{\gamma}}

\newcommand{\omegastar}{\ensuremath{\omegaparam^{*}}}
\newcommand{\gammahat}{\ensuremath{\hat{\gammaparam}}}
\newcommand{\gammatilde}{\ensuremath{\gammaparam^{*}}}

\newcommand{\Fset}{\ensuremath{F}}
\newcommand{\Cset}{\ensuremath{C}}
\newcommand{\Aset}{\ensuremath{A}}
\newcommand{\AsetC}{\ensuremath{A^{\complement}}}
\newcommand{\Bset}{\ensuremath{B}}
\newcommand{\BsetOmega}{\ensuremath{B_{\omegaparam}}}
\newcommand{\BsetOmegaC}{\ensuremath{B_{\overline{\omegaparam}}}}
\newcommand{\bOmegaC}{\ensuremath{b}}
\newcommand{\bOmegaPrime}{\ensuremath{b'}}
\newcommand{\bOmegaT}{\ensuremath{b_{\overline{\omegaparam}}(t)}}

\newcommand{\nFUnknown}{\ensuremath{N_{1}}}
\newcommand{\nBUnknown}{\ensuremath{N_{2}}}

\newcommand{\LHSfunc}{\ensuremath{\mathcal{L}}}
\newcommand{\LHSfuncparam}{\ensuremath{\LHSfunc(\gammaparam, \omegaparam)}}
\newcommand{\LHSfuncprime}{\ensuremath{\LHSfunc(\gammaparam', \omegaparam')}}

\newcommand{\kgamma}{\ensuremath{k_1}}
\newcommand{\komega}{\ensuremath{k_2}}
\newcommand{\kk}{\ensuremath{k_3}}

\newcommand{\wplusd}{\ensuremath{w^{+}(d_1, d_2)}}
\newcommand{\wid}{\ensuremath{w_{i}^{*}(d_i)}}

\newcommand{\wtwo}{\ensuremath{w_{2}^{*}(d_2)}}
\newcommand{\wone}{\ensuremath{w_{1}^{*}(d_1)}}

\newcommand{\depstwo}{\ensuremath{d_2/\epsilon_2}}
\newcommand{\depsone}{\ensuremath{d_1/\epsilon_1}}

\newcommand{\epsot}{\ensuremath{\epsilon_{12}}}

\newcommand{\donedaggall}{\ensuremath{\epsilon_1 (2\epsot - 1)/\epsot}}
\newcommand{\doneddaggall}{\ensuremath{\epsilon_1\epsot}}

\newlength\figureheight 
\newlength\figurewidth 

\newtheorem{myLemma}{Lemma}

\newcounter{cnt}

\newcommand{\Prob}{\ensuremath{\operatorname{Pr}}}

\usepackage{caption}
\usepackage[caption=false,font=footnotesize]{subfig}
\usepackage{url}

\begin{document}
%
\title{Two-terminal Erasure Source-Broadcast with Feedback}

\author{\IEEEauthorblockN{Louis Tan, Kaveh Mahdaviani and Ashish Khisti~\IEEEmembership{Member,~IEEE}}
\thanks{L.~Tan, K.~Mahdaviani and A.~Khisti are with the Dept.\ of Electrical and Computer Engineering, University of Toronto, Toronto, ON, Canada (e-mail: louis.tan@mail.utoronto.ca, mahdaviani@cs.toronto.edu, akhisti@ece.utoronto.ca). 
Part of this work was presented at the 2015 International Symposium on Information Theory in Hong Kong~\cite{TMKS_ISIT15}.}%
}%
\maketitle

\begin{abstract}
%

We study the effects of introducing a feedback channel in the two-receiver erasure source-broadcast problem in which a binary equiprobable source is to be sent over an erasure broadcast channel to two receivers subject to erasure distortion constraints.  The receivers each require a certain fraction of a source sequence, and we are interested in the minimum latency, or transmission time, required to serve them all.  We first show that for a two-user broadcast channel, a point-to-point outer bound can always be achieved.  We further show that the point-to-point outer bound can also be achieved if only one of the users, the \emph{stronger} user, has a feedback channel.  Our coding scheme relies on a hybrid approach that combines transmitting both random linear combinations of source symbols as well as a retransmission strategy.

\end{abstract}



%
\IEEEpeerreviewmaketitle

\section{Introduction\label{sec:intro}}

The presence of a feedback channel can have many benefits.  In addition to practical issues such as reducing complexity, it can also increase fundamental communication rates in multiuser networks~(see, e.g.,~\cite{OzarowCheong84,Cover_TIT98}).  In particular, feedback has been shown to increase the \emph{channel capacity} of the erasure broadcast channel~\cite{GGT,WangHan14}.  In contrast, we study a \emph{joint source-channel coding} problem of broadcasting an equiprobable binary source over an erasure broadcast channel with feedback.  Each receiver demands a certain fraction of the source, and we look to minimize the overall transmission time needed to satisfy all user demands.

We are motivated by heterogeneous broadcast networks, where we encounter users with very different channel qualities, processing abilities, mobility, screen resolutions, etc. In such networks, the user diversity can translate to different distortion requirements from the broadcaster since, e.g., a high-quality reconstruction of a video may not be needed by a user with a  limited-capability device.  In addition, if the source we wish to reconstruct is the output of a multiple description code, then the fraction of the source that is recoverable is of interest.

A related problem has been studied for the case without feedback in~\cite{TLKS_TIT16}.  There are also many other variations of the problem~(see~\cite{LTDM} for a thorough literature review).  A channel coding version of the problem was studied in~\cite{GGT}, which proposed a general algorithm for sending a \emph{fixed} group of messages to $n$ users over an erasure broadcast channel with feedback.  In contrast, our formulation allows flexibility in \emph{which} messages are received at a user so long as the total number received exceeds a certain threshold.  The variant of the index coding problem of \cite{BF_ISIT13} is similar in this respect in that given $n$ users, each already possessing a different subset of $m$ messages, the goal is to minimize the number of transmissions over a \emph{noiseless} channel before \emph{each} user receives \emph{any} additional $t$ messages.  

In our work, we utilize uncoded transmissions that are instantly-decodable~\cite{SorourValaee15}, and distortion-innovative.  The zero latency in decoding uncoded packets has benefits in areas in which packets are instantly useful at their destination such as applications in video streaming and disseminating commands to sensors and robots~\cite{SorourValaee15,SorourValaee10}.  We show that when feedback is available from both users, we can always send instantly-decodable, distortion-innovative transmissions.  While this may not necessarily be the case if a feedback channel is available from only the stronger user, we will show that in this case, the optimal minmax latency can still be achieved by using repetition coding  in tandem with the transmission of random linear combinations. 
\section{{System Model}}
\label{sec:system_model_feedback}






We first consider a version of the erasure-source broadcast problem involving universal feedback, after which, we consider the case of one-sided feedback in Section~\ref{sec:one_sided_feedback}. Our problem involves communicating a binary memoryless source $\{S(t)\}_{t=1,2, \ldots}$ to two users over an erasure broadcast channel with feedback.  
%
%
The source produces equiprobable symbols in $\mathcal{S}=\{0,1\}$ and is communicated by an encoding function that produces the channel input sequence $X^{W} = (X(1),  \dots , X(W))$, where $X(t)$ denotes the $t^{\mathrm{th}}$ channel input taken from the alphabet $\mathcal{X} = \{0, 1\}$.  We assume that 
$X(t)$ is a function of the source as well as the channel outputs of all users prior to time $t$.  


Let $Y_{i}(t)$ be the channel output observed by user $i$ on the $t^{\mathrm{th}}$ channel use for $i \in \{1, 2\}$. 
We let $Y_{i}(t)$ take on values in the alphabet $\mathcal{Y} = \{0, 1, \star\}$ so that an erasure event is represented by `$\star$'.   
For $W \in \mathbb{N}$,  let $[W]$ denote the set $\{1, 2, \ldots, W\}$. We associate user~$i$ with the state sequence $\{Z_i(t)\}_{t \in [W]}$, which represents the noise on user~$i$'s channel, where $Z_i(t) \in \mathcal{Z} \triangleq \{0,1\}$,  and  $Y_i(t)$ will be erased if $Z_{i}(t) = 1$ and $Y_{i}(t) = X(t)$ if $Z_i(t) = 0$.  The channel we consider is memoryless in the sense that $(Z_{1}(t), Z_2(t))$ is drawn i.i.d.\ from the probability mass function given by   

\begin{subequations}
\label{eq:pmf_z1z2}
\begin{align}
	\textrm{Pr}(Z_1 = 1, Z_2 = 1) &= \epsot \\
	\textrm{Pr}(Z_1 = 1, Z_2 = 0) &= \epsilon_1 - \epsot \\
	\textrm{Pr}(Z_1 = 0, Z_2 = 1) &= \epsilon_2 - \epsot \\
	\textrm{Pr}(Z_1 = 0, Z_2 = 0) &= 1 - \epsilon_1 - \epsilon_2 + \epsot,			
\end{align}
\end{subequations}
where $\epsot\in (0, 1)$ is the probability that an erasure simultaneously occurs on both channels and $\epsilon_i \in (0, 1)$ denotes the erasure rate of the channel corresponding to user~$i$, where we assume $\epsilon_1 < \epsilon_2$.

The problem we consider involves causal feedback  that is universally available.  That is, at time $T$, we assume that $\{Z_1(t), Z_2(t)\}_{t=1, 2, \ldots, T-1}$ is available to the transmitter and all receivers. 
After $W$ channel uses, user $i$ utilizes the feedback and his own channel output to reconstruct the source as a length-$N$ sequence, denoted as $\hat{S}_{i}^{N}$.  We will be interested in a fractional recovery requirement so that each symbol in $\hat{S}_{i}^{N}$ either faithfully recovers the corresponding symbol in $S^{N}$, or otherwise a failure is indicated with an erasure symbol, i.e., we do not allow for any bit flips.

More precisely, we choose the reconstruction alphabet $\mathcal{\hat{S}}$ to be an augmented version of the source alphabet so that $\mathcal{\hat{S}} = \{0, 1, \star\}$, where the additional `$\star$' symbol indicates an erasure symbol.  Let $\mathcal{D} = [0,1]$ and $d_i \in \mathcal{D}$ be the distortion user $i$ requires.  We then express the constraint that an achievable code ensures that each user $i \in \{1, 2\}$ achieves a fractional recovery of $1 - d_{i}$ with the following definition.

\begin{mydef}
\label{def:code_two_users_feedback}
	An $(N, W, d_{1}, d_{2})$ code for source $S$ on the erasure broadcast channel with \emph{universal} feedback consists of	
	\begin{enumerate}
		\item a sequence of encoding functions $f_{t, N} : \mathcal{S}^{N} \times  \prod_{j = 1}^{2} \mathcal{Z}^{t-1} \to \mathcal{X}$ for $t \in [W]$, such that $X(t) = f_{t, N}(S^{N}, Z_1^{t -1}, Z_2^{t -1})$, and
		
		\item two decoding functions $g_{i,N} : \mathcal{Y}^{W} \times \mathcal{Z}^{2W} \to \mathcal{\hat{S}}^{N}$ s.t.\ for $i \in \{1, 2\}$, $\hat{S}_{i}^{N} = g_{i,N}(Y_{i}^{W}, Z_1^{W}, Z_2^{W})$, and
		\begin{enumerate}
			\item $\hat{S}_{i}^{N}$ is such that for $t \in [N]$, if $\hat{S}_{i}(t) \neq S(t)$, then $\hat{S}_{i}(t) = \star$,
			\item $\mathbb{E}   \left\vert{\{t \in [N] \mid \hat{S}_{i}(t) = \star\}}\right\vert \leq N d_{i}$.
		\end{enumerate}
		 	
	\end{enumerate}
	
\end{mydef}

We again mention that in our problem formulation, we assume that all receivers have causal knowledge of $(Z_{1}^{t-1}, Z_{2}^{t-1})$ at time $t$.  That is, each receiver has causal knowledge of which packets were received.  This can be made possible, for example, through the control plane of a network.

We define the {\it latency} that a given code requires before all users can recover their desired fraction of the source as follows.
\begin{mydef}
\label{def:latency_two_users}
	The latency, $w$, of an~$(N, W, d_{1}, d_{2})$ code is the number of channel uses per source symbol that the code requires to meet all distortion demands, i.e., $w = W/N$.
\end{mydef}
Our goal is to characterize the achievable latencies under a prescribed distortion vector,
as per the following definition.
\begin{mydef}
\label{def:achievable_general_feedback}
	Latency $w$ is said to be $(d_{1}, d_{2})$-achievable over the erasure broadcast channel if for every $\delta > 0$, there exists for sufficiently large $N$, an $(N, wN, \hat{d}_{1}, \hat{d}_{2})$ code such that for all $i \in \{1, 2\}$, $d_{i}+\delta \geq \hat{d}_{i}$.
	
	

\end{mydef}

\begin{remark}
	We remark that while our definitions have assumed binary source and channel input symbols for simplicity, our results can be easily extended to non-binary alphabets.
\end{remark}

In the next section, we show that the point-to-point Shannon bound can always be achieved for the problem we have just defined.  In Section~\ref{sec:one_sided_feedback}, we present a variation of the problem where only the stronger user has access to a feedback channel.  We further study the case of three receivers in the sequel.  

\section{Source-broadcast with Universal Feedback}
\label{sec:two_users} 
We first show that the case involving only two users can be fully solved.  We do this by demonstrating an algorithm that achieves point-to-point optimality for both users at any time during transmission.  Specifically, for $i \in \{1, 2\}$, let $\wid$ be the point-to-point optimal latency for user~$i$ obtained from the source-channel separation theorem where

\begin{align}
\label{eq:wid_optimal}
	\wid = \frac{1 - d_i}{1 - \epsilon_i}.
\end{align}
We now present an algorithm to achieve this outer bound.  In the first phase of the algorithm, we successively transmit each source symbol uncoded until at least one user receives it. If $S(t)$ is received only by user~$i$,  then the transmitter places $S(t)$ into queue $Q_j$.
No action is taken if both users receive $S(t)$. By assumption, feedback is universally available, and so user~$i$ is also able to maintain a local version of queue~$Q_j$.

Now, after this first phase, the transmitter has built queues $Q_1$ and $Q_2$, where 
for $i,j \in \{1, 2\}, i\neq j$, 
user~$i$ has knowledge of packets in $Q_j$ and is in need of those in $Q_i$.
Thus, the algorithm's next phase involves successively transmitting a linear combination of the packets at the fronts of $Q_1$ and $Q_2$.  Let $q_i$ be the packet at the front of $Q_i$.  Notice that a successfully received channel symbol of the form $q_1 \oplus q_2$ means that user~1 is able to decode $q_1 \in Q_1$,  since he has access to $q_2 \in Q_2$. 
We therefore remove $q_i$ from $Q_i$ whenever a linear combination involving $q_i$ is received by user~$i$.
This entire phase continues until the users' distortion constraints are met.  The decoding algorithm for this scheme is also simple.  Given that user~$i$ has decoded source symbol~$S(t)$ from the linear combination he received, user~$i$ sets $\hat{S}_i(t) = S(t)$.



Our algorithm has two appealing properties. The first is that it involves only transmissions that are \emph{instantly decodable}, which is seldom the case when channel codes are used.  
Secondly, this coding scheme involves only transmissions that are \emph{distortion-innovative}.  This means that any successfully received channel symbol can be immediately used to reconstruct a single source symbol that was hitherto unknown.  
In fact, our coding scheme has the property that for any latency $w \in [0, 1/(1 - \epsilon)]$, after $wN$ transmissions have been sent over the channel, an expected value of $\gamma = wN(1 - \epsilon)$ channel symbols were received, which leads to the decoding of precisely $\gamma$ source symbols.  The distortion achieved after $wN$ transmissions is thus seen to be $D = 1 - w(1 - \epsilon)$, which we readily recognize as 
the separation-based outer bound in~\eqref{eq:wid_optimal}.  Since the transmission of an instantly-decodable, distortion-innovative symbol does not require a channel encoder, we will sometimes refer to such transmissions as \emph{analog} transmissions.

In the next section, we study
a variation of our problem formulation in which only the \emph{stronger} user has a feedback channel available.  For this problem variation, we show that we can still achieve the optimal minmax latency despite the weaker user not having access to a feedback channel.

\section{Source-broadcast with One-sided Feedback}
\label{sec:one_sided_feedback}

In this section we consider a one-sided-feedback variation of the problem in Section~\ref{sec:two_users} whereby in a broadcast network with two receivers, a feedback channel is available to only the stronger user.  In this scenario, we show that given the distortion constraints of both users, there is no overhead in the minmax latency achieved.  Specifically, let $\wplusd$ be the Shannon lower bound for the minmax latency problem we consider (c.f.\ Definition~\ref{def:achievable_general_feedback}) where

\begin{align}
\label{eq:wplusd}
	\wplusd = \max_{i \in \{1, 2\}} \wid,
\end{align}
and $\wid$ is defined in~\eqref{eq:wid_optimal}.
Section~\ref{sec:two_users} showed that for $i \in \{1, 2\}$, user~$i$ can achieve distortion $d_i$ at the optimal latency $\wid$ when a feedback channel is available to \emph{both} users.  Clearly, the optimal minmax system latency $\wplusd$ is also achievable in this case.  In contrast, in this section we show that when a feedback channel is available to \emph{only the stronger user}, while the \emph{individual} optimal latencies may or may not be achievable, the overall system latency $\wplusd$ is still achievable.


\subsection{Problem Formulation}
\label{sec:system_model_one_sided}







\begin{figure}
	\centering
	\includegraphics[scale=1]{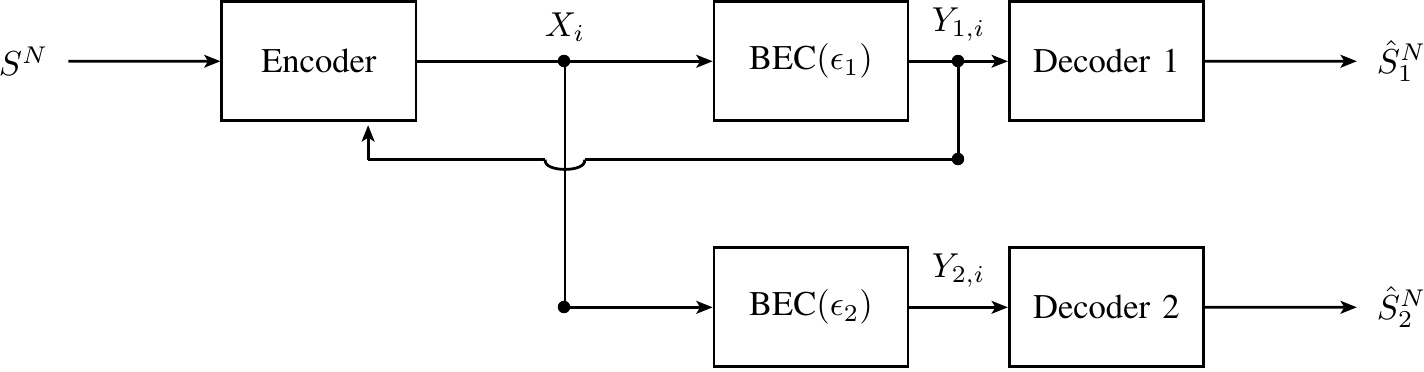}
	\caption{Broadcasting an equiprobable binary source over an erasure broadcast channel with a feedback channel asymmetrically available to only the stronger user.}
	\label{fig:system_model_one_sided}
\end{figure}

The problem is illustrated in Figure~\ref{fig:system_model_one_sided}.  We now consider a problem involving \emph{one-sided} feedback, which is a variation of the problem defined in Section~\ref{sec:system_model_feedback} with the modification that out of the two receivers in the broadcast network, only the receiver with the lower erasure rate has a feedback channel available.  That is, at time $T$, we assume that only $\{Z_1(t)\}_{t=1, 2, \ldots, T-1}$ is available to the transmitter and both receivers rather than having $\{Z_1(t), Z_2(t)\}_{t=1, 2, \ldots, T-1}$ available.  We therefore modify the definition of a code given in Definition~\ref{def:code_two_users_feedback} to suit our current problem involving the erasure source-broadcast problem with \emph{one-sided} feedback with the following definition.

%

\begin{mydef}
\label{def:code_one_sided}
	An $(N, W, d_{1}, d_{2})$ code for source $S$ on the erasure broadcast channel with \emph{one-sided} feedback consists of	
	\begin{enumerate}
		\item A sequence of encoding functions $f_{i, N} : \mathcal{S}^{N} \times  \mathcal{Z}^{i-1} \to \mathcal{X}$ for $i \in [W]$, s.t.\ $X(i) = f_{i, N}(S^{N}, Z_1^{i -1})$
		
		\item Two decoding functions $g_{i,N} : \mathcal{Y}^{W} \times \mathcal{Z}^{W} \to \mathcal{\hat{S}}^{N}$ s.t.\ $\hat{S}_{i}^{N} = g_{i,N}(Y_{i}^{W}, Z_1^{W})$, $i \in \{1, 2\}$, and
		\begin{enumerate}
			\item $\hat{S}_{i}^{N}$ is such that for $t \in [N]$, if $\hat{S}_{i}(t) \neq S(t)$, then $\hat{S}_{i}(t) = \star$,
			\item $\mathbb{E}   \left\vert{\{t \in [N] \mid \hat{S}_{i}(t) = \star\}}\right\vert \leq N d_{i}$.
		\end{enumerate}
		 	
	\end{enumerate}
	
\end{mydef}

The definitions of the \emph{latency} of a code and the \emph{achievable latency region} for the one-sided feedback problem follow Definitions~\ref{def:latency_two_users} and~\ref{def:achievable_general_feedback} respectively and are left out for brevity.  

\subsection{Coding Scheme}

We begin by reviewing a repetition coding scheme in the next subsection, whereby we simply ignore the weaker user, and focus on using the stronger user's feedback to retransmit each of his required source symbols until it is received.  A repetition coding scheme is useful insofar as it helps avoid compelling the weaker user to decode additional source symbols that he does not require.  

As an example, consider when the stronger user requires $N$ source symbols.  We could send random linear combinations of the $N$ symbols, which he could decode after receiving $N$ equations through, say, $W$ transmissions.  By the time the stronger user has recovered the $N$ equations, the weaker user, having a weaker channel, would have received less than $N$ equations.  At this point, the weaker user could simply ignore the first $W$ transmissions, and have the transmitter encode another random linear combination of symbols for the weaker user to decode.  However, such a timesharing scheme is inefficient.  
On the other hand, the weaker user could prevent the first $W$ transmissions from going to waste by continuing to receive random linear combinations of the group of $N$ source symbols originally intended for the stronger user.  However, if the weaker user requires $M < N$ symbols, he would have had to listen to many more transmissions than necessary to recover $M$ symbols thus introducing delay.

We notice the problem is in the random linear combinations used in our coding scheme.  In such a scheme, we either receive more than $N$ equations and decode the entirety of the $N$ source symbols, or we receive less than $N$ equations and decode none of the source symbols.  This ``threshold effect'' is detrimental when we have heterogeneous users in a network who require $M < N$ source symbols.  



The repetition coding scheme avoids this pitfall by avoiding random linear combinations altogether and instead transmitting uncoded source symbols over the channel.  While this avoids compelling the weaker user to decode unnecessary source symbols, it can also be inefficient for the weaker user.  Specifically, since the repetition scheme is based solely on the stronger user's feedback, a source symbol can be retransmitted even after it is received by the weaker user.  We show how to circumvent this problem by creating a hybrid coding scheme that consists of both repetitions, and random linear combinations.  The coding scheme is controlled by two variables, $\omegaparam$, and $\gammaparam$.  We show how to choose specific values for these parameters to achieve the optimal minmax latency in Section~\ref{sec:one_sided_optimality}.


\subsubsection{Repetition Coding}
\label{subsubsec:repetition_coding}

Consider a coding scheme that simply ignores the weaker user and focuses on using the stronger user's feedback to retransmit each of his required source symbols until it is received.  We wish to calculate the expected value of $\Nretransmit$, which we define as the number of \emph{unique} source symbols received by the weaker user when $k \in \mathbb{N}$ symbols are to be sent to the stronger user via repetition coding.

Let $\nRepi{i}$ be a random variable representing the number of transmissions needed to be sent for symbol $S(i)$ to be received by the \emph{stronger} user, user~1, in the repetition scheme.  Let $\indi{i}$ be an indicator variable indicating whether symbol $S(i)$ was received by the \emph{weaker} user, user~2, in any of the $\nRepi{i}$ transmissions and let $\vecnRep = \{\nRepi{1}, \nRepi{2}, \ldots, \nRepi{k}\}$ be the vector of random variables giving the number of repetitions needed to send each source symbol.  
Given $\vecnRep$, we calculate $\mathbb{E}(\Nretransmit | \vecnRep)$ as 

\setcounter{cnt}{1}
\begin{align}
	\mathbb{E}(\Nretransmit | \vecnRep) &= \sum_{i = 1}^{k} \mathbb{E}(\indi{i} | \vecnRep) \\
	&= \sum_{i = 1}^{k} \Prob(\text{user~2 receives } S(i) | \vecnRep) \\
	&\stackrel{(\alph{cnt})}{=} \sum_{i = 1}^{k} \left\{1 - \textrm{Pr}(Z_2 = 1 | Z_{1} = 1)^{\nRepi{i} - 1}\textrm{Pr}(Z_2 = 1 | Z_{1} = 0) \right\}\\
	\addtocounter{cnt}{1}	
	\label{eq:Nretransmit_g_Mi}	
	&\stackrel{(\alph{cnt})}{=} \sum_{i = 1}^{k} \left\{1 - \left( \frac{\epsot}{\epsilon_1} \right)^{\nRepi{i} - 1} \left(\frac{\epsilon_2 - \epsot}{1 - \epsilon_1}\right)\right\},
\end{align}
where 
\begin{enumerate}[(a)]
	\item follows by construction of the repetition-based scheme
	\item we have calculated the conditional probabilities from~\eqref{eq:pmf_z1z2}.
\end{enumerate}

\setcounter{cnt}{1}
We next use the law of total probability to get that 
\begin{align}
	\ENretransmit &= \sum_{\vecnRep} \mathbb{E}(\Nretransmit | \vecnRep) \Prob(\vecnRep) \\
	&\stackrel{(\alph{cnt})}{=} \sum_{\vecnRep} \sum_{i = 1}^{k} \left\{1 - \left( \frac{\epsot}{\epsilon_1} \right)^{\nRepi{i} - 1} \left(\frac{\epsilon_2 - \epsot}{1 - \epsilon_1}\right)\right\}  \Prob(\vecnRep) \\
	\addtocounter{cnt}{1}
	&\stackrel{(\alph{cnt})}{=} \sum_{i = 1}^{k}  \sum_{j= 1}^{\infty}  \left\{1 - \left( \frac{\epsot}{\epsilon_1} \right)^{\nRepi{i} - 1} \left(\frac{\epsilon_2 - \epsot}{1 - \epsilon_1}\right)\right\} \Prob(\nRepi{i} = j)  \\
	\addtocounter{cnt}{1}
	&= k - \left(\frac{\epsilon_2 - \epsot}{1 - \epsilon_1}\right)\sum_{i = 1}^{k}  \sum_{j= 1}^{\infty} \left( \frac{\epsot}{\epsilon_1} \right)^{\nRepi{i} - 1}  \Prob(\nRepi{i} = j)  \\
	&\stackrel{(\alph{cnt})}{=} k - \left(\frac{\epsilon_2 - \epsot}{1 - \epsilon_1}\right) \sum_{i = 1}^{k}  \sum_{j= 1}^{\infty} \left( \frac{\epsot}{\epsilon_1} \right)^{j - 1} (\epsilon_{1}^{j - 1}(1 - \epsilon_1))  \\
	\addtocounter{cnt}{1}	
	&= k \left( 1 - (\epsilon_2 - \epsot)  \sum_{j= 1}^{\infty} (\epsot)^{j-1}   \right) \\
	&\stackrel{(\alph{cnt})}{=} k \left( 1 - (\epsilon_2 - \epsot) \left\{ \frac{1}{1 - \epsot}\right\}   \right) \\
	\label{eq:ENretransmit}
	&=  k \cdot \frac{1 - \epsilon_2}{1 -\epsot},
\end{align}
where 
\begin{enumerate}[(a)]
	\item follows from~\eqref{eq:Nretransmit_g_Mi}
	\item follows from the fact that the $\nRepi{i}$ are i.i.d.
	\item follows from the fact that by construction, $\nRepi{i}$ is a geometric random variable with probability of success $(1 - \epsilon_1)$
	\item follows from the formula for the geometric series and the fact that $\epsot < 1$.
\end{enumerate}

\begin{myLemma}
\label{lem:repetition}
	Let $k$ source symbols be sent to the stronger user via a repetition scheme.  Then $\ENretransmit $, the expected number of unique source symbols received by the weaker user, is given by~\eqref{eq:ENretransmit}.
\end{myLemma}
\subsubsection{Inner Bound}
\label{subsubsec:inner_bound}

In this section, we formulate a hybrid coding scheme that incorporates both repetition coding and sending random linear combinations of source symbols.  The code is tuned via two parameters, $\omegaparam, \gammaparam \in [0, 1]$.  We target point-to-point optimal performance for the stronger user in this section, and show that this is possible for any values of $\omegaparam$ and $\gammaparam$.  In the next section however, we show how to optimize $\omegaparam$ and $\gammaparam$ to achieve the optimal minmax latency.  As in the coding scheme in Section~\ref{sec:two_users}, we again split the coding scheme into phases.  

In Phase~\Rmnum{1}, we begin by sending each source symbol uncoded over the channel.  That is, Phase~\Rmnum{1} consists of $N$ transmissions and at time $t \in \{1, 2, \ldots, N\}$, we transmit $X(t) = S(t)$.  Let $\Aset \subseteq \{S(1), S(2), \ldots, S(N)\}$ be the set of symbols received by the stronger user in Phase~\Rmnum{1}.  Since the stronger user's feedback is available to all receivers and the transmitter, $\Aset$ is known to all parties.  

\begin{figure}
	\centering
	\includegraphics[scale=1]{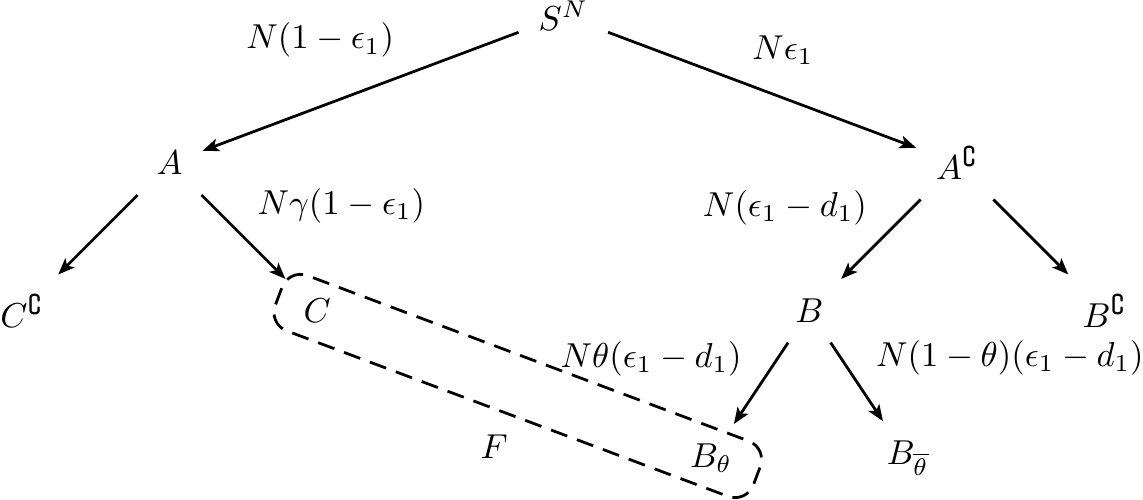}
	\caption{A tree diagram illustrating the relationship between the sets of source symbols.  Each node represents a set of source symbols and a directed edge $(X,Y)$ indicates that set $Y$ is a subset of $X$.  If edge $(X, Y)$ is also \emph{weighted}, then the weight represents the expected cardinality of set $Y$.  Only sets involved in the coding scheme have incoming \emph{weighted} edges.  The direct successors of a node form a partition for the set representing the parent node.  The root of the tree is the entire source sequence, which is subsequently partitioned at each depth of the tree.  We also show set $\Fset$, which is the union of $\Cset$ and $\BsetOmega$.}
	\label{fig:set_construction}
\end{figure}

At the conclusion of Phase~\Rmnum{1}, we have that on average, for $i \in \{1, 2\}$, user~$i$ will have received $N(1 - \epsilon_i)$ source symbols and so will require an additional $N(\epsilon_i - d_i)$ symbols in the remaining phases.   Before moving on to Phase~\Rmnum{2}, we first organize the source symbols in $\Aset$ and $\AsetC$ into subsets, where $\AsetC \subseteq \{S(1), S(2), \ldots, S(N)\}$ denotes the complement of set $\Aset$. We first isolate a fraction of $N(\epsilon_1 - d_1)$ source symbols from $\AsetC$ into a set denoted as $\Bset$.  That is, we fix the remaining $N(\epsilon_1 - d_1)$ symbols that the stronger user requires in $\Bset$.  We then partition $\Bset$ into two disjoint sets, one that contains a fraction of $\omegaparam \in [0, 1]$ source symbols from $\Bset$, denoted as $\BsetOmega$, and the other that contains the remaining fraction of $1 - \omegaparam$ symbols, denoted as $\BsetOmegaC$, where $\Bset = \BsetOmega \cup \BsetOmegaC$.  Random linear combinations of the symbols in $\BsetOmega$ will be sent to the stronger user while the symbols in $\BsetOmegaC$ will be sent with repetition coding.  We further take a fraction of $\gammaparam \in [0, 1]$ source symbols from $\Aset$ and denote this set as $\Cset$.  Finally, we define $\Fset$ as the union of sets $\Cset$ and $\BsetOmega$.  Figure~\ref{fig:set_construction} illustrates the relationship between all sets and the manner in which they are constructed.

In Phase~\Rmnum{2} of the coding scheme, we designate the symbols of $\BsetOmegaC$ as the symbols to be transmitted to the stronger user with a repetition scheme.  However, we modify the repetition scheme to incorporate random linear combinations of symbols in $\Fset$.  In a conventional repetition scheme, we would retransmit $\bOmegaC \in \BsetOmegaC$ until it is received by the stronger user.  Upon reception by the stronger user, we move on to the next symbol $\bOmegaC' \in \BsetOmegaC$ and continue in this manner until all symbols in $\BsetOmegaC$ are accounted for.  Let $\bOmegaT \in \BsetOmegaC$ be the source symbol being repeated at time $t$.  Our modified coding scheme is similar to the conventional repetition scheme except that at any time $t$, instead of \emph{only} transmitting $\bOmegaT$, we instead send $v(t) + \bOmegaT$, where $v(t)$ is a \emph{new} random linear combination of the source symbols in $\Fset$ generated for every time $t$.  Let  $\bOmegaC \in \BsetOmegaC$.  If $\bOmegaT = \bOmegaC$ is transmitted and subsequently received by the stronger user at time $t$, the protocol for replacing $\bOmegaC$ at time $t+ 1$ with another source symbol from $\BsetOmegaC$ is identical to the conventional repetition scheme, however the only difference is that we now combine $\bOmegaT$ with a random linear combination of the symbols of $\Fset$ at every transmission.  Phase~\Rmnum{2} concludes when all symbols in $\BsetOmegaC$ have been accounted for by the modified repetition scheme.

\begin{remark}
	When applying a random linear code, we use the maximum distance separable (MDS)-type property that any collection of $N$ channel symbols gives $N$ linearly independent equations.  Although strictly speaking such codes do not exist over the binary field, randomly chosen combinations over long blocks are approximately MDS~\cite{TTAAJ17}.
\end{remark}

At the conclusion of Phase~\Rmnum{2}, since we have transmitted the symbols in $\BsetOmegaC$ as if we were utilizing a repetition scheme, we have that on average, the stronger user will have received $|\BsetOmegaC|$ equations involving $|\BsetOmegaC| + |\Fset|$ variables.  Notice, however, that since $\Fset \triangleq \Cset \cup \BsetOmega$, and $\Cset \subseteq \Aset$, the stronger user can subtract off all symbols originating from $\Cset$.  Therefore, Phase~\Rmnum{1} actually results in the stronger user receiving $|\BsetOmegaC|$ equations involving $|\BsetOmegaC| + |\BsetOmega| = |\Bset|$ \emph{unkown} variables, where $\mathbb{E}|\Bset| = N(\epsilon_1 - d_1)$.  The stronger user therefore requires an additional $|\BsetOmega|$ equations at the conclusion of Phase~\Rmnum{2}.

In Phase~\Rmnum{3}, we send the remaining equations to the stronger user by continuing to send $v(t)$ at any time $t$.  That is, we continue sending random linear combinations of the symbols in $\Fset$.  Phase~\Rmnum{3} concludes when the feedback of the stronger user indicates that he has received the missing $|\BsetOmega|$ equations.

At the conclusion of Phase~\Rmnum{3}, it is not hard to see that the stronger user achieves point-to-point optimal performance, since every channel symbol received has provided an independent equation that can be used to decode a new source symbol.  
At this point, if $\wtwo \leq \wone$, we halt any further transmissions, where $\wid$ is given by~\eqref{eq:wid_optimal}.

In Phase~\Rmnum{4}, if $\wtwo > \wone$, we continue to transmit $v(t)$, the random linear combinations of the source symbols in $\Fset$, for an additional $N(\wtwo- \wone)$ transmissions.

\subsection{Minmax Latency Optimality}
\label{sec:one_sided_optimality}

In this section, we show that it is possible to choose values for $\omegaparam, \gammaparam \in [0, 1]$ from Section~\ref{subsubsec:inner_bound} so that the lower bound for the minmax latency in~\eqref{eq:wplusd} is achieved.  We first calculate the expected number of \emph{unknown} variables involved in transmissions to the weaker user from Phase~\Rmnum{2} onwards.  

First, since we send random linear combinations of the symbols in $\Fset$ in Phase~\Rmnum{2}, we initially expect this to contribute $|\Fset|$ variables.  However, some of the symbols in $\Fset$ have already been received by user~2 in Phase~\Rmnum{1}.  Let $\nFUnknown$ be the number of symbols in $\Fset$ \emph{not} received by user~2 in Phase~\Rmnum{1}.  Given a channel noise realization $(Z_{1}^{W}, Z_{2}^{W}) = (z_{1}^{W}, z_{2}^{W})$, we can calculate the expected value of $\nFUnknown$ as

\setcounter{cnt}{1}
\begin{align}
	\mathbb{E}(\nFUnknown | (Z_{1}^{W}, Z_{2}^{W}) = (z_{1}^{W}, z_{2}^{W})) &= \sum_{s \in \Fset} \textrm{Pr}(\textrm{$s$ not received by user~2 in Phase~\Rmnum{1}})\\
	&\stackrel{(\alph{cnt})}{=} \sum_{s \in \Cset} \textrm{Pr}(\textrm{$s$ not received by user~2 in Phase~\Rmnum{1}}) \\ \nonumber
	&\qquad  + \sum_{s' \in \BsetOmega} \textrm{Pr}(\textrm{$s'$ not received by user~2 in Phase~\Rmnum{1}})\\ 
	\addtocounter{cnt}{1}
	&\stackrel{(\alph{cnt})}{=} \sum_{s \in \Cset} \textrm{Pr}(Z_2 = 1 | Z_1 = 0) + \sum_{s' \in \BsetOmega} \textrm{Pr}(Z_2 = 1 | Z_1 = 1) \\ 
	\addtocounter{cnt}{1}
	&\stackrel{(\alph{cnt})}{=} \sum_{s \in \Cset} \left(\frac{\epsilon_2 - \epsot}{1 - \epsilon_1}\right) + \sum_{s' \in \BsetOmega} \left( \frac{\epsot}{\epsilon_1}\right) \\ 
	\addtocounter{cnt}{1}
	\label{eq:last_line_set}
	&= |\Cset| \left(\frac{\epsilon_2 - \epsot}{1 - \epsilon_1} \right) + |\BsetOmega|\left( \frac{\epsot}{\epsilon_1}\right),
\end{align}
where 
\begin{enumerate}[(a)]
	\item follows from the fact that $\Fset = \Cset \cup \BsetOmega$ and $\Cset$ and $\BsetOmega$ are disjoint by construction
	\item follows from the fact that by construction, all symbols in $\Cset$ have been received by user~1 and all symbols in $\BsetOmega$ were not received by user~1
	\item we have calculated the conditional probabilities from~\eqref{eq:pmf_z1z2}.
\end{enumerate}

The cardinality of sets $\Cset$ and $\BsetOmega$ depends on the channel noise variables $(Z_{1}^{W}, Z_{2}^{W})$.  By taking the expectation over the channel noise, we can calculate the unconditional expected value of $\nFUnknown$ as

\setcounter{cnt}{1}
\begin{align}
	\label{eq:nFUnknown}
	\mathbb{E}\nFUnknown &\stackrel{(\alph{cnt})}{=} N \gammaparam (1 - \epsilon_1) \left(\frac{\epsilon_2 - \epsot}{1 - \epsilon_1} \right) + N \omegaparam(\epsilon_1 - d_1) \left( \frac{\epsot}{\epsilon_1}\right),
\end{align}
where 
\begin{enumerate}[(a)]
	\item follows from~\eqref{eq:last_line_set} and by construction of the sets (see Section~\ref{subsubsec:inner_bound} and Figure~\ref{fig:set_construction}).
\end{enumerate}

%

The use of repetition coding for the symbols in $\BsetOmegaC$ in Phase~\Rmnum{2} further adds additional unknown variables to the coding scheme.  On average, the expected number of symbols repeated is $\mathbb{E}|\BsetOmegaC| = N(1  - \omegaparam)(\epsilon_1 - d_1)$, of which, again, only a fraction of $\textrm{Pr}(Z_2 = 1 | Z_1 = 1) = \epsot/\epsilon_1$ were not already received by the weaker user in Phase~\Rmnum{1}.  By Lemma~\ref{lem:repetition}, the number of additional \emph{unknown} variables introduced to the weaker user as a result of the repetition scheme is therefore given by $\nBUnknown$, where

\setcounter{cnt}{1}
\begin{align}
	\label{eq:nBUnknown}
	\mathbb{E}\nBUnknown &= N(1  - \omegaparam)(\epsilon_1 - d_1)\left(\frac{\epsot}{\epsilon_1}\right) \left(\frac{1 - \epsilon_2}{1 - \epsot} \right).
\end{align}

Let $\LHSfuncparam$ be the expected fraction of all source symbols that are involved in transmissions to the weaker user from Phase~\Rmnum{2} onwards that have not yet been decoded prior to Phase~\Rmnum{2}.  We have that $\LHSfuncparam$ is the normalized sum of~\eqref{eq:nFUnknown} and~\eqref{eq:nBUnknown}, i.e., 

\begin{align}
	\LHSfuncparam &= \frac{\mathbb{E}\nFUnknown + \mathbb{E}\nBUnknown}{N} \\
	&= \kgamma \gammaparam + \komega \omegaparam	+ \kk,
	\label{eq:LHSfuncparam}
\end{align}
where 
\begin{subequations}
\begin{align}
	\kgamma &= \epsilon_2 - \epsot, \label{eq:kgamma} \\	
	\komega &= (\epsilon_1 - d_1) \left(\frac{\epsot}{\epsilon_1}\right) \left(\frac{\epsilon_2 - \epsot}{1 - \epsot} \right), 	\label{eq:komega}\\
	\kk &= (\epsilon_1 - d_1) \left(\frac{\epsot}{\epsilon_1}\right) \left(\frac{1 - \epsilon_2}{1 - \epsot} \right).
	\label{eq:kk}
\end{align}
\end{subequations}

Having calculated the number of unknown variables sent to the weaker user from Phase~\Rmnum{2} onwards, we now consider the number of equations he receives in Phases~\Rmnum{2} and~\Rmnum{3}.  From Section~\ref{subsubsec:inner_bound}, we know that during these phases, the total number of transmissions was simply equal to the number of trials needed to send $N(\epsilon_1 - d_1)$ equations to the stronger user with feedback.  The number of transmissions in Phases~\Rmnum{2} through~\Rmnum{3}  is therefore distributed according to a negative binomial distribution and the mean number of transmissions in this period is $W_{2,3} = N(\epsilon_1 - d_1)/(1 -\epsilon_1)$.  Of these transmissions, the expected number received by the weaker user is equal to $W_{2, 3}(1 - \epsilon_2)$.  We rewrite the expression for $W_{2, 3}(1 - \epsilon_2)$, the expected number of transmissions received by user~2 in Phases~\Rmnum{2} through~\Rmnum{3}, as $NC_{2, 3}$ where

\begin{align}
	C_{2,3} = \frac{(\epsilon_1 - d_1)(1 - \epsilon_2)}{1 - \epsilon_1}.
\end{align}
We next compare $N\LHSfuncparam$, the amount of source symbols destined for the weaker user, with $NC_{2, 3}$,  the expected number of equations received over the weaker user's channel during Phases~\Rmnum{2} and~\Rmnum{3}.  

As mentioned in Section~\ref{subsubsec:inner_bound}, the weaker user requires an additional $N(\epsilon_2 - d_2)$ symbols to be sent from Phase~\Rmnum{2} onwards.  Therefore, it is necessary that $\LHSfuncparam \geq \epsilon_2 - d_2$.  However, if $\LHSfuncparam$ is much greater than $\epsilon_2 - d_2$, we encounter the problem explained in the introduction of this section in which the weaker user is forced to decode unnecessary symbols thus introducing delay.  Say that we are able to find values of $\gammaparam', \omegaparam' \in [0, 1]$ such that $\LHSfuncprime = \epsilon_2 - d_2$.  We consider two cases when this is so -- when $\LHSfuncprime \leq C_{2, 3}$ and when $\LHSfuncprime > C_{2, 3}$.  We show that in both cases, we can achieve the optimal minmax  latency so long as $\LHSfuncprime = \epsilon_2 - d_2$.

In the first case, when $\LHSfuncprime \leq C_{2, 3}$, we wish to send less information over the channel than what the channel can support.  Therefore, we expect that the weaker user should be able to decode all source symbols before the conclusion of Phase~\Rmnum{3}.  However, in general, if the weaker user achieves distortion $d_2$ after decoding, it will be at a latency $w$, where $w > \wtwo$.  That is, in general, the weaker user may not achieve an \emph{individual} point-to-point optimal latency.

To see why this is so, recall from Section~\ref{subsubsec:inner_bound} that in Phase~\Rmnum{2} of our coding scheme, we transmit $\bOmegaT + v(t)$ at every time instant $t$, where $v(t)$ is a new random linear combination of the source symbols in $\Fset$ generated at every time $t$.  Since $\LHSfuncprime \leq C_{2, 3}$, there is the possibility that at some point, the weaker user is able to decode all symbols in $\Fset$ even before Phase~\Rmnum{2} has concluded.  Say that this is the case and the stronger user has stalled on receiving a particular symbol $\bOmegaC \in \BsetOmegaC$ being repeated.  Let us further assume that the weaker user has already received $\bOmegaC$.  Then while $\bOmegaC + v(t)$ is being transmitted, all transmissions to the weaker user are redundant.
After $\bOmegaC$ is received by the stronger user and the transmitter moves on to  $\bOmegaPrime$, the next symbol in $\BsetOmegaC$ to be sent via the modified repetition scheme, the weaker user can continue to receive innovative information.  However, the set of transmissions received while $\bOmegaC$ is being repeated prevents the weaker user from achieving an optimal \emph{individual} latency.

However, we show that the optimal \emph{minmax} latency can still be achieved.  Notice that the moment all symbols in $\Fset$ can be decoded by the weaker user, the random linear combination $v(t)$ can be subtracted from any transmission $\bOmegaT + v(t)$ in Phase~\Rmnum{2}.  Therefore the remainder of Phase~\Rmnum{2} effectively consists of uncoded transmissions from the weaker user's perspective, and he is eventually able to decode all $\LHSfuncprime$ symbols.  Thus, so long as $\LHSfuncprime = \epsilon_2 - d_2$, the weaker user will decode the necessary amount of symbols before the conclusion of Phase~\Rmnum{3}, while the stronger user decodes at an optimal latency the moment Phase~\Rmnum{3} terminates.  In this case, the stronger user is the bottleneck of the system and in fact, the condition $\LHSfuncprime \leq C_{2, 3}$ is equivalent to $\wone\geq \wtwo$.

On the other hand, if $\LHSfuncprime > C_{2, 3}$, the weaker user has more unknown variables than equations and so he cannot yet decode at the conclusion of Phase~\Rmnum{3}.  However, every transmission he has received so far is ``innovative'' in the sense that it provides an independent equation that can be used to decode the $N\LHSfuncprime$ source symbols.  In order to decode, we simply need to send additional equations to the weaker user, and since $\LHSfuncprime = \epsilon_2 - d_2$, there will not be any unnecessary source symbols sent.  Since the stronger user is point-to-point optimal at the conclusion of Phase~\Rmnum{3}, we have therefore sent a total of $N\wone$ transmissions up to that point.  Since, from the weaker user's perspective, we have hitherto been sending random linear combinations of $N\LHSfuncprime$ variables, we simply need to continue doing so for another $N(\wtwo - \wone)$ transmissions in Phase~\Rmnum{4} before he receives the remaining number of equations required and achieves point-to-point optimal performance.

\begin{table}
	\begin{center}
		\begin{tabular}{| c | c |}
			\hline
			\multicolumn{2}{|c|}{{\bf Ordering of Boundaries for $d_1$}} \\
			\hline
			{\bf Inequality} & {\bf Justification}   \\ \hline
			$\donedaggall < \doneddaggall$ & $(1 - \epsot)^2 > 0$ \\ \hline 
			$\doneddaggall < \epsilon_1$ & $\epsot < 1$, $\epsilon_1 > 0$ \\ \hline 
			\hline
		\end{tabular}
	\end{center}
	\caption{We justify the ordering of the region boundaries for $d_1$.  In the left column, we have the ordering between two boundary points, and in the right column, we show the necessary and sufficient condition that justifies the ordering.}	
	\label{tab:d1_boundaries}	
\end{table}

We therefore see that regardless of whether $\LHSfuncparam$ is greater or less than $C_{2, 3}$, we can achieve an optimal minmax latency so long as we can find $\gammaparam, \omegaparam \in [0, 1]$ such that $\LHSfuncparam = \epsilon_2 - d_2$.   We focus on finding these values of $\gammaparam$ and $\omegaparam$ in the next sections. In doing so, we consider three cases cases for $d_1$.  
%
The first is when $0 \leq d_1 \leq \donedaggall$, the second when $\donedaggall < d_1 < \doneddaggall$, and third when $\doneddaggall \leq d_1 < \epsilon_1$.  We justify the position of these boundary points with Table~\ref{tab:d1_boundaries}.  For example, in the first row of Table~\ref{tab:d1_boundaries}, we justify that the boundary point $\donedaggall$ is less than the boundary point $\doneddaggall$ with the necessary and sufficient condition that $(1 - \epsot)^2 > 0$.  

After dividing the values of $d_1$ into regions, we then further consider regions of $d_2/\epsilon_2$, where each region requires a distinct choice for the values of $\gammaparam$ and $\omegaparam$.  We note that from Remark~1 in~\cite{TLKS_TIT16} that for $i \in \{1, 2\}$, we assume that $d_i/\epsilon_i < 1$, otherwise an uncoded transmission strategy can achieve~\eqref{eq:wplusd}.  Therefore, we consider only values of $d_2/\epsilon_2 \in [0, 1]$.  In the following sections, the regions of $\depstwo$ will depend on the boundaries $\cstar$, $\mydstar$, $\astar$ and $\bstar$, which we define as


\begin{align}
	\cstar &= \left(\frac{\epsot}{\epsilon_2}\right) \left(\frac{d_1}{\epsilon_1}\right), \\
	\mydstar &= \frac{d_1\epsot + \epsilon_1(\epsilon_2 - \epsot)}{\epsilon_1\epsilon_2}, \\	
	\astar &= \frac{\epsot (d_1 (1 - \epsilon_2) + \epsilon_1(\epsilon_2 - \epsot))}{(1 - \epsot)\epsilon_1 \epsilon_2},\\
	\bstar &= \frac{d_1\epsot(1 - \epsilon_2) + \epsilon_1(\epsilon_2 - \epsot)}{(1 - \epsot) \epsilon_1 \epsilon_2}.
\end{align}


\subsubsection{Case~\Rmnum{1}: $0 \leq d_1 \leq \donedaggall$}


\begin{figure}
	\centering
	\includegraphics[scale=1]{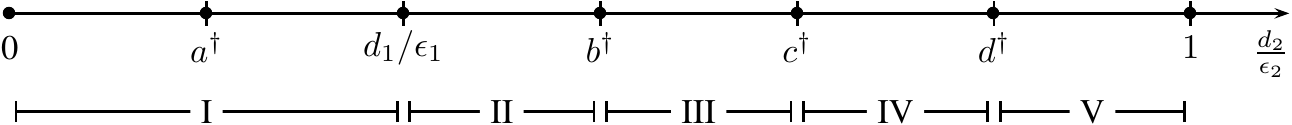}
	\caption{The different regions requiring separate coding schemes when $0 \leq d_1 \leq \donedaggall$.}
	\label{fig:regions_d1_small}
\end{figure}

\begin{table}
	\begin{center}
		\begin{tabular}{| c | c |}
			\hline
			\multicolumn{2}{|c|}{{\bf Ordering of Boundaries for $\depstwo$ when $0 \leq d_1 \leq \donedaggall$}} \\
			\hline
			{\bf Inequality} & {\bf Justification}   \\ \hline
			$\cstar < d_1/\epsilon_1$ & $\epsot < \epsilon_2$ \\ \hline 
			$d_1/\epsilon_1 < \mydstar $ & $d_1 < \epsilon_1$ \\ \hline 
			$\mydstar \leq \astar$ & $d_1 \leq \donedaggall$ \\ \hline 
			$\astar < \bstar$ & $\epsot < 1$ \\ \hline 
			$\bstar < 1$ & $d_1 < \epsilon_1$ \\
			\hline
		\end{tabular}
	\end{center}
	\caption{We justify the ordering of the region boundaries of Figure~\ref{fig:regions_d1_small} when $0 \leq d_1 \leq \donedaggall$.  In the left column, we have the ordering between two boundary points, and in the right column, we show the necessary and sufficient condition that justifies the ordering.}	
	\label{tab:d1_small}
\end{table}

We first mention that the upper bound in the assumption $d_1 \leq \donedaggall$ can be either positive or negative depending on the values of $\epsilon_1$ and $\epsot$.  In the case that the upper bound is positive and $d_1$ is in this region, we now go through the process of finding the values of $\gammaparam$ and $\omegaparam$ such that $\LHSfuncparam = \epsilon_2 - d_2$.  


We begin by dividing the number line for $\depstwo$ in Figure~\ref{fig:regions_d1_small}, where we have justified the ordering of each boundary point with Table~\ref{tab:d1_small}.  For example, in the third row of the table, we justify the ordering that $\mydstar \leq \astar$ with the necessary and sufficient condition that $d_1 \leq \donedaggall$, which is the assumption in this region of $d_1$ that we consider.  We enumerate all regions of Figure~\ref{fig:regions_d1_small} and provide the values of $\gammaparam$ and $\omegaparam$.


\begin{LaTeXdescription}
	\item[Region~\Rmnum{5}] In this region, we set $\gammaparam = \omegaparam = 0$ and recover the unmodified repetition coding scheme discussed in the beginning of Section~\ref{subsubsec:repetition_coding}.  In fact, we do not require that $\LHSfuncparam = \epsilon_2 - d_2$ in this region.  Instead, by repeating all $N(\epsilon_1 - d_1)$ source symbols, by Lemma~\ref{lem:repetition}, we can work out that $N\kk$ uncoded source symbols are received by the weaker user, where $\kk$ is given by~\eqref{eq:kk}.  We can confirm that in Region~\Rmnum{5}, in which $\depstwo \geq \bstar$, the distortion requirement of the weaker user is sufficiently large so that it can be met by the amount of uncoded symbols he receives.
	\item[Region~\Rmnum{4}] In this region, we set $\omegaparam = 0$, and $\gammaparam = \gammatilde$ where 
		\begin{align}
			\gammatilde &= \frac{\epsilon_1(\epsilon_2 - d_2)(1 - \epsot) - \epsot(\epsilon_1 - d_1)(1 - \epsilon_2)}{\epsilon_1(1 - \epsot)(\epsilon_2 - \epsot)}.
			\label{eq:gammatilde}
		\end{align}
		Within this region, the weaker user's distortion constraint is sufficiently low so that additional coded symbols must be transmitted.  We can confirm that this choice of $\gammaparam$ and $\omegaparam$ results in $\LHSfuncparam = \epsilon_2 - d_2$ in~\eqref{eq:LHSfuncparam}.  Furthermore, the conditions $\gammatilde \geq 0$ and $\gammatilde \leq 1$ are equivalent to $\depstwo \leq \bstar$ and $\depstwo \geq \astar$ respectively, which are satisfied in this region.
	\item [Region~\Rmnum{3}]  In this region, we set $\gammaparam = 0$ and $\omegaparam = \omegastar$, where
		\begin{align}
			\omegastar &= \frac{\epsilon_1(\epsilon_2 - d_2)(1 - \epsot) - \epsot(\epsilon_1 - d_1)(1 - \epsilon_2)}{\epsot(\epsilon_1 - d_1)(\epsilon_2 - \epsot)}.
		\end{align}
		We can confirm that this choice of $\gammaparam$ and $\omegaparam$ results in $\LHSfuncparam = \epsilon_2 - d_2$ in~\eqref{eq:LHSfuncparam}.  Furthermore, the conditions $\omegastar \geq 0$ and $\omegastar \leq 1$ are equivalent to $\depstwo \leq \bstar$ and $\depstwo \geq \mydstar$ respectively, which are satisfied in this region.
	
	\item[Region~\Rmnum{2}]  In this region, we set $\omegaparam= 1$, $\gammaparam = \gammahat$, where
		\begin{align}
			\gammahat &= \frac{\epsilon_1(\epsilon_2 - d_2) - \epsot(\epsilon_1 - d_1)}{\epsilon_1(\epsilon_2 - \epsot)}.
			\label{eq:gammahat}
		\end{align}
		We can confirm that this choice of $\gammaparam$ and $\omegaparam$ results in $\LHSfuncparam = \epsilon_2 - d_2$ in~\eqref{eq:LHSfuncparam}.  Furthermore, the conditions $\gammahat \geq 0$ and $\gammahat \leq 1$ are equivalent to $\depstwo \leq \mydstar$ and $\depstwo \geq \cstar$ respectively, which are satisfied in this region.
	
	\item[Region~\Rmnum{1}]  In this region, we ignore feedback altogether and use the successive segmentation coding scheme of~\cite{TLKS_TIT16}, which showed that both users can be point-to-point optimal if $\depstwo \leq \depsone$.
\end{LaTeXdescription}


\subsubsection{Case~\Rmnum{2}: $\donedaggall < d_1 < \doneddaggall$}

\begin{figure}
	\centering
	\includegraphics[scale=1]{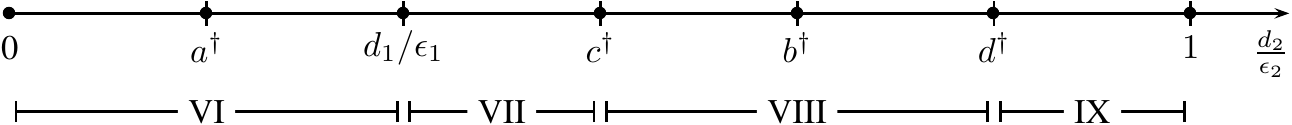}
	\caption{The different regions requiring separate coding schemes when $\donedaggall < d_1 < \doneddaggall$.}
	\label{fig:regions_d1_mid}
\end{figure}

We again divide the number line for $\depstwo$ in Figure~\ref{fig:regions_d1_mid} where we have justified the ordering of each boundary point with Table~\ref{tab:d1_mid}.  Notice however, that in the ranges of $d_1$ we now consider, the boundary points $\astar$ and $\mydstar$ have swapped positions compared to Figure~\ref{fig:regions_d1_small}.  We again enumerate all regions of Figure~\ref{fig:regions_d1_mid}.

\begin{table}
	\begin{center}
		\begin{tabular}{| c | c |}
			\hline
			\multicolumn{2}{|c|}{{\bf Ordering of Boundaries for $\depstwo$ when $\donedaggall < d_1 < \doneddaggall$}} \\
			\hline
			{\bf Inequality} & {\bf Justification}   \\ \hline
			$\cstar < d_1/\epsilon_1$ & $\epsot < \epsilon_1$ \\ \hline 
			$d_1/\epsilon_1 < \astar $ & $d_1 < \doneddaggall$ \\ \hline 
			$\astar < \mydstar $ & $d_1 > \donedaggall$ \\ \hline 			
			$\mydstar < \bstar$ & $d_1 < \epsilon_1$ \\ \hline 
			$\bstar < 1$ & $d_1 < \epsilon_1$ \\
			\hline
		\end{tabular}
	\end{center}
	\caption{We justify the ordering of the region boundaries of Figure~\ref{fig:regions_d1_mid} when $\donedaggall < d_1 < \doneddaggall$.  In the left column, we have the ordering between two boundary points, and in the right column, we show the necessary and sufficient condition that justifies the ordering.}	
	\label{tab:d1_mid}	
\end{table}

\begin{LaTeXdescription}
	\item[Region~\Rmnum{9}] In this region, we set $\gammaparam = \omegaparam = 0$ and recover the unmodified repetition coding scheme discussed in the beginning of Section~\ref{subsubsec:repetition_coding} (see the description of Region~\Rmnum{5} in the previous section).

	\item[Region~\Rmnum{8}] In this region, we set $\omegaparam = 0$, and $\gammaparam = \gammatilde$ where $\gammatilde$ is given by~\eqref{eq:gammatilde} (see the description of Region~\Rmnum{4} in the previous section).
	
	\item[Region~\Rmnum{7}]  In this region, we set $\omegaparam= 1$, $\gammaparam = \gammahat$, where $\gammahat$ is given by~\eqref{eq:gammahat} (see the description of Region~\Rmnum{2} in the previous section).

	\item[Region~\Rmnum{6}]  In this region, we again ignore feedback altogether and use the successive segmentation coding scheme of the previous chapter (see the description of Region~\Rmnum{1} in the previous section).
\end{LaTeXdescription}

\subsubsection{Case~\Rmnum{3}: $\doneddaggall \leq d_1 < \epsilon_1$}

\begin{figure}
	\centering
	\includegraphics[scale=1]{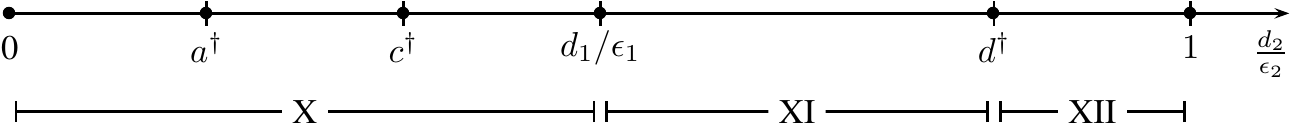}
	\caption{The different regions requiring separate coding schemes when $\doneddaggall \leq d_1 < \epsilon_1$.}
	\label{fig:regions_d1_big}
\end{figure}

For the final case of $d_1$ we consider, we again divide the number line for $\depstwo$ in Figure~\ref{fig:regions_d1_big}, where we have justified the ordering of each boundary point with Table~\ref{tab:d1_big}.  Again, the values of $d_1$ we consider have resulted in some of the boundaries in Figure~\ref{fig:regions_d1_big} moving relative to Figure~\ref{fig:regions_d1_mid}, most notably that now $\astar < \depsone$.  We again enumerate all regions of Figure~\ref{fig:regions_d1_mid}.

\begin{table}
	\begin{center}
		\begin{tabular}{| c | c |}
			\hline
			\multicolumn{2}{|c|}{{\bf Ordering of Boundaries for $\depstwo$ when $\doneddaggall \leq d_1 < \epsilon_1$}} \\
			\hline
			{\bf Inequality} & {\bf Justification}   \\ \hline
			$0 < \cstar$ & $d_1 > 0$ \\ \hline 
			$\cstar < \astar$ & $d_1 < \epsilon_1$ \\ \hline 
			$\astar < d_1/\epsilon_1$ & $d_1 \geq \doneddaggall$ \\ \hline 
			$d_1/\epsilon_1 < \bstar$ &$d_1 < \epsilon_1$ \\ \hline 
			$\bstar < 1$ & $d_1 < \epsilon_1$ \\
			\hline
		\end{tabular}
	\end{center}
	\caption{We justify the ordering of the region boundaries of Figure~\ref{fig:regions_d1_big} when $\doneddaggall \leq d_1 < \epsilon_1$.  In the left column, we have the ordering between two boundary points, and in the right column, we show the necessary and sufficient condition that justifies the ordering.}	
	\label{tab:d1_big}	
\end{table}

\begin{LaTeXdescription}
	\item[Region~\Rmnum{12}] In this region, we set $\gammaparam = \omegaparam = 0$ and recover the unmodified repetition coding scheme discussed in the beginning of Section~\ref{subsubsec:repetition_coding} (see the description of Region~\Rmnum{5} in the previous section).

	\item[Region~\Rmnum{11}] In this region, we set $\omegaparam = 0$, and $\gammaparam = \gammatilde$ where $\gammatilde$ is given by~\eqref{eq:gammatilde} (see the description of Region~\Rmnum{4} in the previous section).

	\item[Region~\Rmnum{10}]  In this region, we again ignore feedback altogether and use the successive segmentation coding scheme of the previous chapter (see the description of Region~\Rmnum{1} in the previous section).
\end{LaTeXdescription}

\section{Conclusion}
\label{sec:conclusion}

In this paper, we considered the erasure source-broadcast problem in which a binary source is to be sent over an erasure broadcast channel with feedback to two receivers with erasure distortion constraints.  To the best of our knowledge, the erasure source-broadcast problem \emph{without} feedback has not been solved in the sense that to date, a set of coinciding inner and outer bounds for the latency region has not been found.  However, when studying the problem \emph{with} feedback, we have derived two conclusive results.  

Our first result was for the case in which causal feedback is universally available from both receivers.  We showed that for this problem, the Shannon source-channel separation bound can be achieved for both receivers.  That is, both receivers can achieve the same performance as if the other receiver was not present.  

We then studied a variation of the previous problem involving one-sided feedback in which causal feedback is available from only the stronger user.  We showed that the Shannon source-channel separation outer bound for the minmax latency can still be achieved despite the fact that we have feedback from only one user.  

\ifCLASSOPTIONcaptionsoff
  \newpage
\fi



%
%
%
%

\vspace{-1em}

\bibliographystyle{IEEEtran}
\bibliography{IEEEabrv,itw2013emina2}

\vfill


\end{document}